\documentclass[twocolumn]{aastex63}
\usepackage{float}

\received{June 1, 2019}
\revised{January 10, 2019}
\accepted{\today}
\submitjournal{ApJL}

\shorttitle{An 8.8 Minute Eclipsing Binary}
\shortauthors{Burdge et al.}
\graphicspath{{./}{figures/}}

\begin{document}

\title{An 8.8 minute orbital period eclipsing detached double white dwarf binary}

\correspondingauthor{Kevin B. Burdge}
\email{kburdge@caltech.edu}

\author[0000-0002-7226-836X]{Kevin B. Burdge}
\affiliation{Division of Physics, Mathematics and Astronomy, California Institute of Technology, Pasadena, CA 91125, USA}

\author[0000-0002-8262-2924]{Michael W. Coughlin}
\affiliation{School of Physics and Astronomy, University of Minnesota, Minneapolis, Minnesota 55455, USA}

\author[0000-0002-4544-0750]{Jim Fuller}
\affiliation{Division of Physics, Mathematics and Astronomy, California Institute of Technology, Pasadena, CA 91125, USA}

\author[0000-0001-6295-2881]{David L. Kaplan}
\affil{Department of Physics, University of Wisconsin-Milwaukee, Milwaukee, WI 53211, USA}

\author[0000-0001-5390-8563]{S. R. Kulkarni}
\affiliation{Division of Physics, Mathematics and Astronomy, California Institute of Technology, Pasadena, CA 91125, USA}

\author[0000-0002-2498-7589]{Thomas R. Marsh}
\affiliation{Department of Physics, University of Warwick, Coventry CV4 7AL, UK}

\author[0000-0002-8850-3627]{Thomas A. Prince}
\affiliation{Division of Physics, Mathematics and Astronomy, California Institute of Technology, Pasadena, CA 91125, USA}

\author[0000-0001-8018-5348]{Eric C. Bellm}
\affiliation{DIRAC Institute, Department of Astronomy, University of Washington, 3910 15th Avenue NE, Seattle, WA 98195, USA}

\author[0000-0002-5884-7867]{Richard G. Dekany}
\affiliation{Caltech Optical Observatories, California Institute of Technology, Pasadena, CA, USA}

\author[0000-0001-5060-8733]{Dmitry A. Duev}
\affiliation{Division of Physics, Mathematics and Astronomy, California Institute of Technology, Pasadena, CA 91125, USA}

\author[0000-0002-3168-0139]{Matthew J. Graham}
\affiliation{Division of Physics, Mathematics and Astronomy, California Institute of Technology, Pasadena, CA 91125, USA}

\author[0000-0003-2242-0244]{Ashish~A.~Mahabal}
\affiliation{Division of Physics, Mathematics and Astronomy, California Institute of Technology, Pasadena, CA 91125, USA}
\affiliation{Center for Data Driven Discovery, California Institute of Technology, Pasadena, CA 91125, USA}

\author[0000-0002-8532-9395]{Frank J. Masci}
\affiliation{IPAC, California Institute of Technology, 1200 E. California
             Blvd, Pasadena, CA 91125, USA}

\author[0000-0003-2451-5482]{Russ R. Laher}
\affiliation{IPAC, California Institute of Technology, 1200 E. California
             Blvd, Pasadena, CA 91125, USA}
             
\author[0000-0002-0387-370X]{Reed Riddle}
\affiliation{Caltech Optical Observatories, California Institute of Technology, Pasadena, CA, USA}
             
\author[0000-0001-6753-1488]{Maayane T. Soumagnac}
\affiliation{Lawrence Berkeley National Laboratory, 1 Cyclotron Road, Berkeley, CA 94720, USA}
\affiliation{Department of Particle Physics and Astrophysics, Weizmann Institute of Science, Rehovot 76100, Israel}




\begin{abstract}

We report the discovery of ZTF J2243+5242, an eclipsing double white dwarf binary with an orbital period of just $8.8$ minutes, the second known eclipsing binary with an orbital period less than ten minutes. The system likely consists of two low-mass white dwarfs, and will merge in approximately 400,000 years to form either an isolated hot subdwarf or an R Coronae Borealis star. Like its $6.91\, \rm min$ counterpart, ZTF J1539+5027, ZTF J2243+5242 will be among the strongest gravitational wave sources detectable by the space-based gravitational-wave detector \emph{The Laser Space Interferometer Antenna} (\emph{LISA}) because its gravitational-wave frequency falls near the peak of \emph {LISA}'s sensitivity. Based on its estimated distance of $d=2120^{+131}_{-115}\,\rm pc$, \emph{LISA} should detect the source within its first few months of operation, and should achieve a signal-to-noise ratio of $87\pm5$ after four years. We find component masses of $M_A= 0.349^{+0.093}_{-0.074}\,M_\odot$ and $M_B=0.384^{+0.114}_{-0.074}\,M_\odot$, radii of $R_A=0.0308^{+0.0026}_{-0.0025}\,R_\odot$ and $R_B = 0.0291^{+0.0032}_{-0.0024}\,R_\odot$, and effective temperatures of $T_A=22200^{+1800}_{-1600}\,\rm K$ and $T_B=16200^{+1200}_{-1000}\,\rm K$. We determined all of these properties, and the distance to this system, using only photometric measurements, demonstrating a feasible way to estimate parameters for the large population of optically faint ($r>21 \,  m_{\rm AB}$) gravitational-wave sources which the Vera Rubin Observatory (VRO) and \emph{LISA} should identify.

\end{abstract}



\section{Introduction} \label{sec:intro}

The population of known double white dwarfs (DWDs) which will merge within a Hubble time (orbital periods $\lesssim 12 \rm \, hrs$) has increased substantially over the last decade, in large part due to efforts such as the extremely low mass white dwarf (ELM) survey \citep{ELMI,ELMII,ELMIII,ELMIV,ELMV,ELMVI,ELMVII,ELMVIII} and the Supernova Type Ia Progenitor (SPY) survey \citep{Napiwotzki2003,Napiwotzki2020}. Over the past two years, through massive expansions in densely sampled time-domain photometric measurements, the Zwicky Transient Facility (ZTF) has facilitated a rapid growth in the population of known DWDs with orbital periods under an hour \citep{Burdge2020}. Two of the sources discovered by ZTF so far, the eclipsing DWD binaries ZTF J1539+5027 ($P_b\approx6.91 \rm \, min$) \citep{Burdge2019a} and ZTF J0538+1953 ($P_b\approx14.44 \rm \, min$) \citep{Burdge2020}, should be detected by \emph{The Laser Space Interferometer Antenna} (\emph{LISA}) \citep{Amaro2017} with a high signal-to-noise ratio, enabling precise parameter estimation using gravitational waves \citep{Littenberg2019}. Thus, using the gravitational wave signal from such a system combined with electromagentic constraints, we will be able to probe novel white dwarf (WD) physics such as the efficiency of tides in these objects \citep{Piro2019}.

Here, we report the discovery of ZTF J2243+5242, a DWD binary with an orbital period of just $8.8 \rm \, min$, the second shortest eclipsing binary system known at the time of discovery. \mbox{ZTF J2243+5242} is a high signal-to-noise (SNR) \emph{LISA}-detectable gravitational-wave source which should be detected within the first month of \emph{LISA}'s operation and reach an SNR of $87\pm5$ four years into the mission. Unique among the binary systems known at $P_b<10 \rm \, min$, this system likely consists of a pair of helium-core white dwarfs (He WDs) or hybrid (helium/CO core) WDs with a mass ratio near unity, suggesting that it will result in a merger \citep{Marsh2004}. This binary is also unique among known $P_b<10 \rm \, min$ systems because neither object is near to filling its Roche lobe (our inferred parameters suggest $\frac{R}{R_L}\approx\frac{2}{3}$ for both objects, where $R$ is the volume-averaged WD radius, and $R_L$ is the radius of the Roche lobe), indicating that the system is well detached. Here, we discuss the properties of this system, its past and future evolutionary history, and prospects for the discovery of more such sources in the eras of \emph{LISA}) and the Vera Rubin Observatory (VRO) \citep{Ivezic2019}.

\section{Observations} \label{sec:obs}

Before we discuss the discovery and analysis of this object (Section \ref{sec:analysis}), we briefly discuss the different data-sets and observations that we used.

\subsection{ZTF Observations}

ZTF is a northern sky synoptic survey based on
observations with the 48-inch Samuel Oschin Schmidt telescope at Palomar Observatory \citep{Bellm2019,Masci2019,Graham2019,Dekany2020}. The camera has a 47 $\rm deg^2$ field of view, and reaches a $5 \sigma$ limiting apparent magnitude of approximately $20.8$ in $g$-band, $20.6$ in $r$-band, and $20.2$ in $i$-band, with standard $30\rm\,s$ exposures.

\begin{figure}
\includegraphics[width=0.5\textwidth]{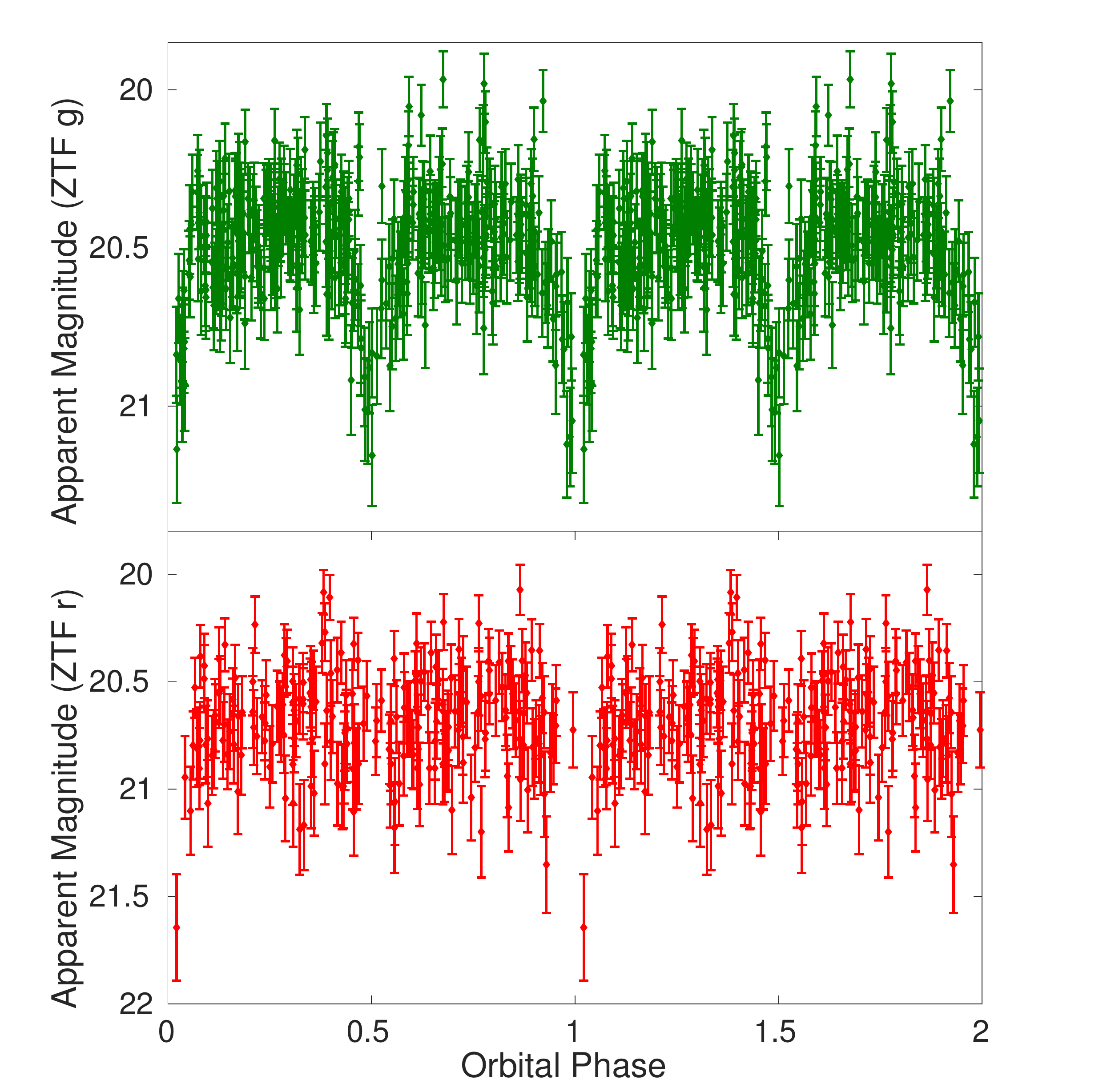}
\caption{Archival ZTF $g$-band (top) and $r$-band (bottom) lightcurves of the system folded at a period of $527.934814\pm0.000021$\,s. 
Because the system is $30\%$ percent brighter in $g$-band than in $r$-band, and ZTF also slightly more sensitive in $g$-band, the discovery was enabled primarily by the $g$-band data.
\label{fig:ZTFLC}}
\end{figure}

\begin{figure}
\includegraphics[width=0.45\textwidth]{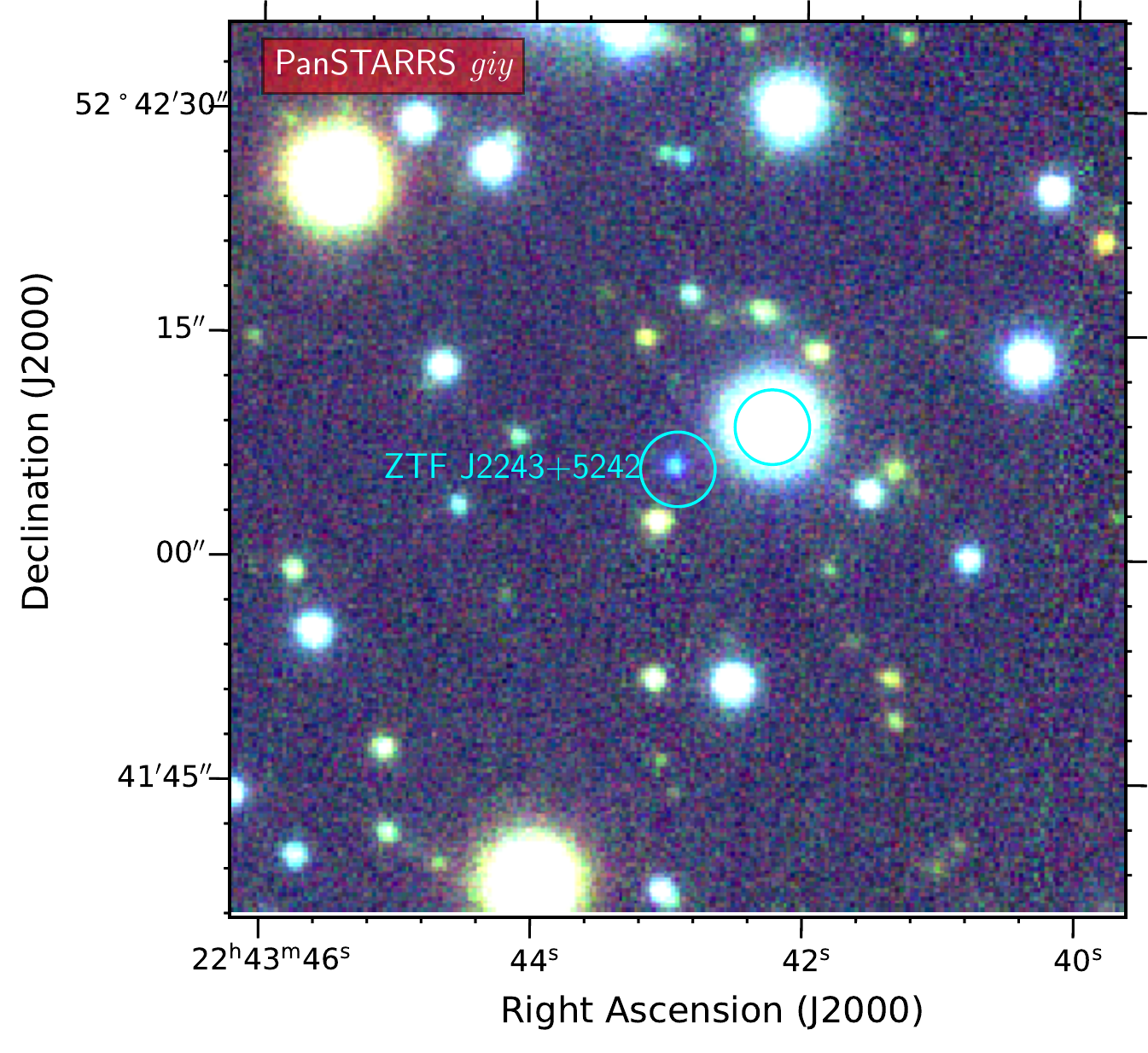}
\caption{A $60\arcsec \times 60\arcsec$ Pan-STARRS1 color $giy$-bands image of ZTF J2243+5242, which is the blue object in the center of the image. We illustrate $2\farcs5$ apertures around the source and a nearby bright star in cyan. Due to the source's proximity to the bright star to the north west (\emph{Gaia} $G\approx 14.5\,m_{\rm V}$), we extracted forced difference image photometry for the ZTF lightcurve we used when modelling the source, PSF photometry for the CHIMERA high speed photometry, and used a $2\farcs5$ aperture for extracting Swift UVOT photometry rather than the default $5\arcsec$ radius.
\label{fig:PS}}
\end{figure}

ZTF J2243+5242 had $218$ $r$-band and $382$ $g$-band good quality photometric detections in its ZTF archival lightcurves at the time of this writing. As illustrated by Figure \ref{fig:ZTFLC}, the discovery was enabled primarily by the $g$-band lightcurve, probably because the object is approximately $30\%$ brighter in $g$-band than it is in $r$-band, and because ZTF is also more sensitive in $g$-band than in $r$-band \citep{Masci2019}. Note that the ZTF archive only contains $5 \sigma$ detections in science images, but in order to model the ZTF lightcurve, after discovery, we extracted forced photometry from ZTF difference images to obtain the best quality lightcurve possible. Using difference images helped improve the photometry significantly due to nearby bright star to the north west, as seen in the Pan-STARRS1 image cutout shown in Figure \ref{fig:PS}. The ZTF lightcurves extracted using forced photometry contained $1384$ $r$-band and $827$ $g$-band observations \citep{Yao2019}.

\subsection{High-speed photometry}

\begin{figure*}
\includegraphics[width=1.0\textwidth]{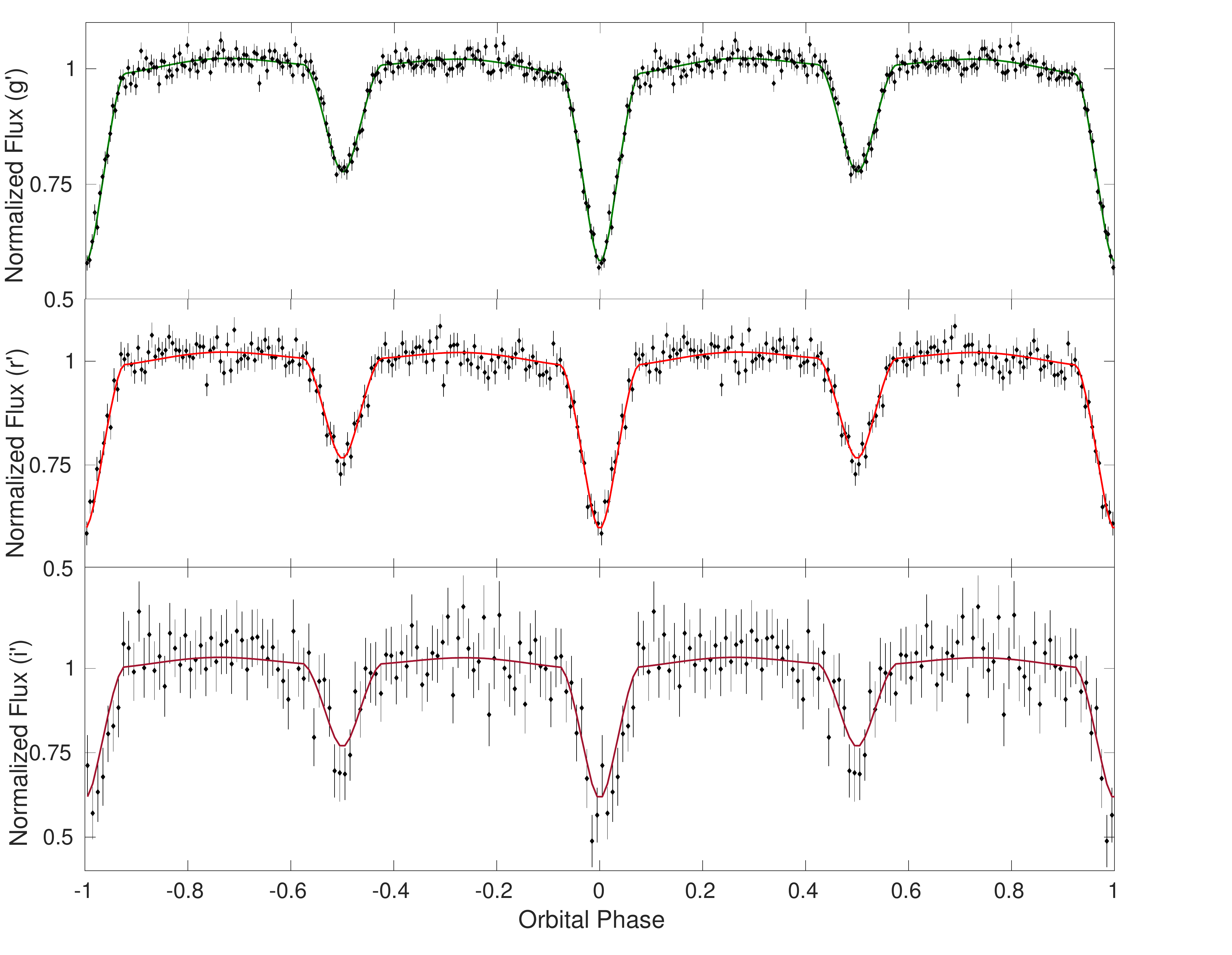}
\caption{The binned, phase-folded CHIMERA $g^\prime$ (top), $r^\prime$ (middle), and $i^\prime$ (bottom) lightcurves of the system, with the best fit LCURVE \citep{Copperwheat2010} model overplotted. 
\label{fig:CHIMERA_LCURVE}}
\end{figure*}

We obtained high speed photometric follow-up of the system using the dual-channel high speed photometer CHIMERA \citep{Harding2016} on the 200-inch Hale telescope at Palomar observatory. We conducted a campaign of observations over several nights, using $g^\prime$ as the blue channel filter, and alternating between $r^\prime$ and $i^\prime$ on the red channel. The phase-folded and binned lightcurves from these observations can be seen in Figure \ref{fig:CHIMERA_LCURVE}. We used a combination of $3\,\rm s$ and $5\,\rm s$ exposure times, due to variable conditions across our nights of observing. All CHIMERA data were reduced using a publicly available pipeline\footnote{\url{https://github.com/mcoughlin/kp84}}, with a newly implemented PSF photometry mode to accommodate reductions for this object, which has a bright neighboring star. Because we were not read-noise limited, we operated the CCD in frame transfer mode using the conventional (as opposed to the electron multiplying) 1 MHz amplifier. On nights of poor seeing ($>1\arcsec$), we binned the readout 2x2 in order to reduce the read-noise. For further details, please see Table \ref{tab:Observations}.

\subsection{Spectroscopic follow-up}

\begin{figure*}
\includegraphics[width=1.0\textwidth]{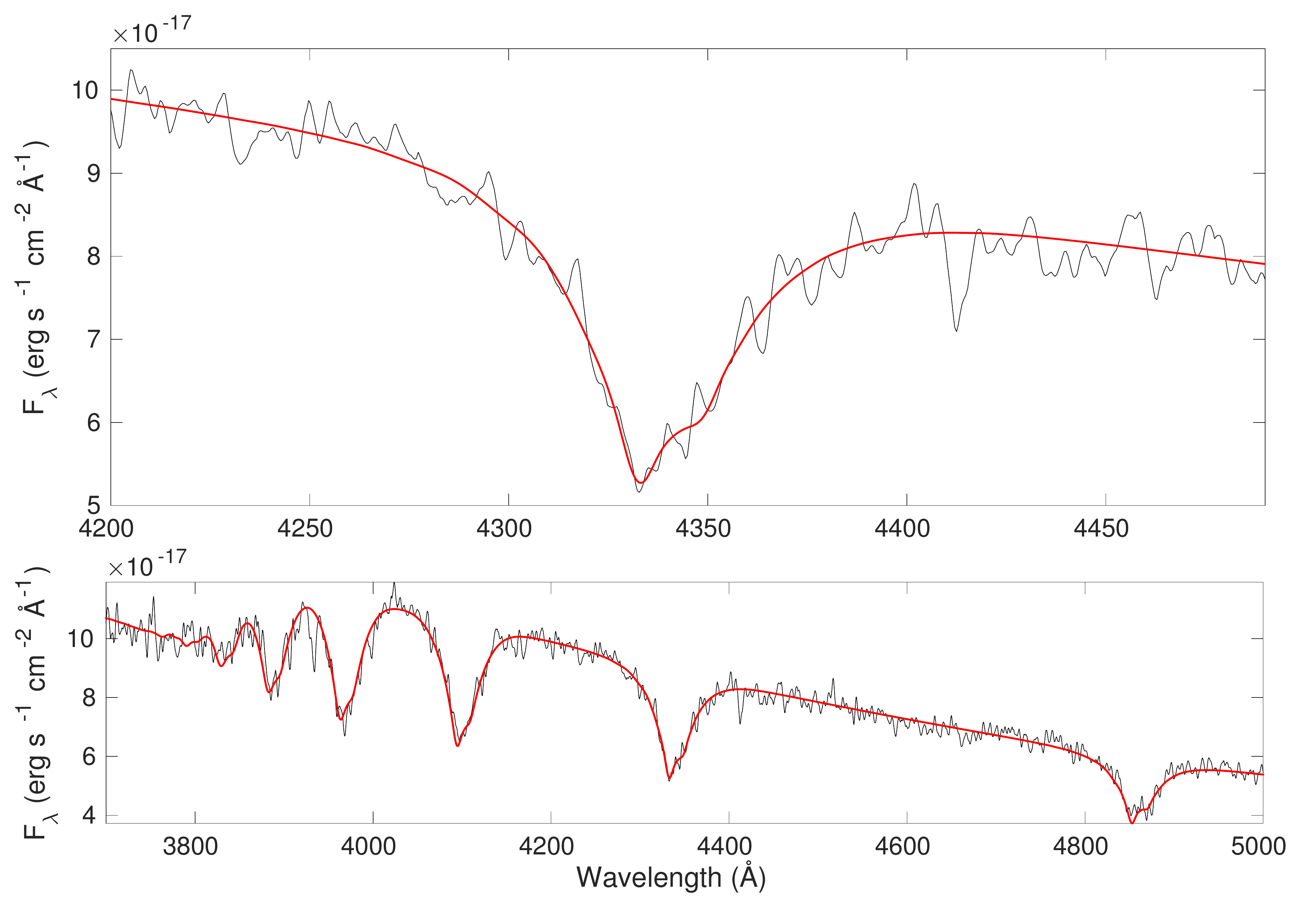}
\caption{An example of a spectroscopic model fit to a phase-binned spectrum of ZTF J2243+5242. Such fits were performed on 12 phase-binned spectra, and used a composite spectrum of two WD models with relative luminosity contributions and effective surface temperatures fixed by lightcurve modelling. The splitting seen in the line cores is indicative that the system is double-lined. We did not measure radial-velocities from these spectra due to the low signal-to-noise ratio.
\label{fig:KeckSpectra}}
\end{figure*}

Using the Low Resolution Imaging Spectrometer (LRIS) on the 10-m W.\ M.\ Keck I Telescope on Mauna Kea \citep{Oke1995}, we conducted phase-resolved spectroscopy on the object. We used an exposure time of $66\,\rm s$, about one eighth of the orbital period, in order to avoid significant Doppler smearing over the course of an exposure. A coadded spectrum of one phase bin is illutrated in Figure \ref{fig:KeckSpectra}. Due to issues with the red channel, we only analyzed data from the blue channel, which covered a wavelength range of approximately $3200 \rm \, \AA$ to $5500 \rm \,\AA$. We used the 600/4000 grism as the dispersive element, and binned the readout 4x4 in order to decrease the readout time to $30\rm \, s$. We obtained a total of 312 exposures (see Table \ref{tab:Observations}). We reduced the data with the publically available lpipe pipeline \citep{Perley2019}, and in order to construct our phase-binned spectra, we divided the orbital phase into 12 bins, and coadded all spectra with a mid-exposure time falling within each bin.

\subsection{Swift observations}

We targeted the binary system with a $5075 \, \rm s$ observation from the \emph{Neil Gehrels Swift Observatory} on 2020~April~02 in order to obtain ultraviolet photometry for the source using the UVOT instrument (see Table \ref{tab:Phot}), as well as an observation with the X-ray telescope (XRT) (Observation ID: 00013301001) \citep{Gehrels2004}. The UVOT observation used 4 exposures with lengths 448\,s--1708\,s, all with the UVM2 filter (centered at 2246\,\AA). 

\begin{deluxetable*}{cccccc}[htbp]
\tablenum{1}
\tablecaption{Table of observations \label{tab:Observations}}
\tablehead{\colhead{Telescope} & \colhead{Instrument} & \colhead{Filter/Mode}& \colhead{Date (UTC)}& \colhead{\# of Exposures}& \colhead{Exposure Time (s)}}
\tablewidth{50pt}
\startdata
  \hline
    Palomar 48-inch & ZTF & ZTF $g$  &  Apr 24 2018--Sep 05 2020 & \phn827  &  \phn\phn30 \\
     Palomar 48-inch & ZTF & ZTF $r$  &  Apr 09 2018--Sep 04 2020 & 1384  &  \phn\phn30 \\
  \hline
     Palomar 200-inch & CHIMERA & $g^\prime$  &  July 15 2020 & 1500  &  \phn\phn\phn5 \\
  \hline
 Palomar 200-inch & CHIMERA & $g^\prime$  &  Jul 21 2020 & 4100  &  \phn\phn\phn5 \\
 Palomar 200-inch & CHIMERA & $r^\prime$  &  Jul 21 2020 & 4100  &  \phn\phn\phn5 \\
    \hline
  Palomar 200-inch & CHIMERA & $g^\prime$  &  Jul 23 2020 & 5000  &  \phn\phn\phn3 \\
 Palomar 200-inch & CHIMERA & $i^\prime$  &  Jul 23 2020 & 2100  &  \phn\phn\phn5 \\
   \hline
 Palomar 200-inch & CHIMERA & $g^\prime$  &  Aug 19 2020 & 6800  &  \phn\phn\phn3 \\
 Palomar 200-inch & CHIMERA & $r^\prime$  &  Aug 19 2020 & 4100  &  \phn\phn\phn5 \\
  \hline
  Keck I & LRIS & Blue Arm  &  Jul 18 2020 & \phn165  & \phn\phn66 \\
  \hline
  Keck I & LRIS & Blue Arm  &  Sept 16 2020 & \phn147  & \phn\phn66 \\
  \hline
  \emph{Swift} & UVOT & UVM2  &  Apr 02 2020 & \phn\phn\phn4  & 5075 \\
  \emph{Swift}  & XRT & PC  &  Apr 02 2020 & \phn\phn\phn1  & 5075 \\
  \hline
\enddata
\end{deluxetable*}

\section{Discovery and analysis} \label{sec:analysis}

\subsection{Photometric selection}

Like the systems described in \citet{Burdge2020}, ZTF J2243+5242 was selected using a broad color cut using Pan-STARRS \citep{Chambers2016} which encompassed all objects with $g-r<0.2$ and $r-i<0.2$ (see \citealt{Burdge2020} for further details). As seen by the apparent magnitudes listed in Table \ref{tab:Phot}, the object's temperature is large enough that it has a color of $g-r\approx-0.21$, and thus could have been targeted with a more restrictive selection. Currently, it is feasible to systematically search a broad selection, but in the VRO era, more restrictive selections may prove valuable in reducing the number of candidates. It is worth noting that the only two binary systems with even shorter orbital periods, HM Cnc \citep{Ramsay2002} and ZTF J1539+5027 \citep{Burdge2019a}, also both exhibit exceptionally blue Pan-STARRS1 colors, of $g-r\approx-0.28$ and $g-r\approx-0.39$, respectively. Unlike HM Cnc and ZTF J1539+5027, which are both substantially brighter in the ultraviolet than in the optical, ZTF J2243+5242 is fainter in these bands, due to modest extinction resulting from its location in the Galactic plane ($b\approx-5.5\degr$).

\begin{deluxetable}{ccl}[htbp]
\tablenum{2}
\tablecaption{Photometric apparent magnitudes and astrometry \label{tab:Phot}}
\tablehead{\colhead{Survey} & \colhead{Filter/Quantity}& \colhead{Measured Value}}
\tablewidth{50pt}
\startdata
\emph{Swift} UVOT& $\rm UVM2$ & $20.73\pm0.10\, m_{\rm AB}$     \\
Pan-STARRS1& $g$ & $20.359\pm0.029\, m_{\rm AB}$     \\
Pan-STARRS1& $r$ & $20.571\pm0.027\, m_{\rm AB}$      \\
Pan-STARRS1& $i$ & $20.733\pm0.024\, m_{\rm AB}$     \\
Pan-STARRS1& $z$ & $20.92\pm0.12\, m_{\rm AB}$    \\
\emph{Gaia}& $G$ & $20.635\pm0.016\, m_{\rm V}$    \\
\emph{Gaia}& RA & $340.929043146\rm\,deg\pm1.05\,\rm mas$     \\
   \emph{Gaia}& Dec & $+52.701660186\rm\,deg\pm0.85\,\rm mas$      \\
   \emph{Gaia}& Parallax & $-1.57\pm1.05\,\rm mas$     \\
   \emph{Gaia}& pm RA & $+0.48\pm2.29\,\rm mas\,yr^{-1}$    \\
   \emph{Gaia}& pm Dec & $-5.12\pm2.10\,\rm mas\,yr^{-1}$    \\
   & E(g$-$r) & $0.16\pm0.02\,m_{\rm AB}$    \\
\hline
\hline
\enddata
\tablecomments{Reddenning estimated using distance reported in Table \ref{tab:Parameters}, with extinction maps of \citet{Green2019}.}
\end{deluxetable}

\subsection{Period finding}

ZTF J2243+5242 was discovered using a graphics processing unit (GPU) based implementation of the conditional entropy algorithm \citep{Graham2013} in the cuvarbase package\footnote{\url{https://github.com/johnh2o2/cuvarbase}}, executed on four Nvidia 2080 Ti GPUs. Notably, because the system exhibits two similar depth eclipses, it was detected at half its period (at $\approx 4.4\,\rm min$), and until we obtained follow-up photometry, it was unclear whether the object had a $4.4\,\rm min$ or $8.8\,\rm min$ orbital period.

\subsection{Swift UVOT and XRT Results}
In the \textit{Swift} UVOT data, we could see ZTF J2243+5242 in the images, but there was a brigher source about $7\farcs5$ to the north west that complicated photometry.  Rather than use the default aperture of $5\arcsec$ radius (where the point spread functions overlap) we measured photometry for  ZTF J2243+5242 using a $2\farcs5$ radius.  We first summed the individual exposures using \texttt{uvotimsum} and then performed aperture photometry with a $2\farcs5$ radius using \texttt{uvotsource}, with a nearby region with radius $40\arcsec$ used to define the background.  We find a source magnitude of $20.73\pm0.10\,m_{\rm AB}$ for ZTF J2243+5242, which has been corrected to $5\arcsec$ radius using the default point spread function present in the \textit{Swift} \texttt{CALDB}.  We include a systematic uncertainty of $0.05\,$mag to account for standard \textit{Swift} processing as well as our non-standard aperture choice. 

For the \textit{Swift} XRT data, there was no obvious emission present at the position of ZTF J2243+5242.  There is 1 event within a circle with radius $9\arcsec$ (the half-power point of the XRT) centered on ZTF J2243+5242.  This was entirely consistent with background emission, where we find a mean of $0.58$\,counts in similar circles randomly distributed across the image.  Therefore we can set a 3-$\sigma$ upper limit of 3\,counts in 5044\,s, or a rate limit of $<0.6\times 10^{-3}\,{\rm count\,s}^{-1}$.

\subsection{Lightcurve+SED modelling and parameter estimation}

We modelled ZTF J2243+5242 by fitting the CHIMERA lightcurve with a model generated using LCURVE \citep{Copperwheat2010}, while simultaneously fitting the Pan-STARRS1 and \textit{Swift} photometry listed in Table \ref{tab:Phot}. Here, we describe this modelling procedure in detail.

Our overall modelling procedure sampled over 14 free parameters: the component masses $M_A$ and $M_B$, temperatures $T_A$ and $T_B$, volume averaged radii $R_A$ and $R_B$, orbital inclination $i$, time of superior conjuction $T_0$, period $P_b$, period derivative $\dot{P}_b$, distance to the system $d$, and three absorption parameters $\alpha_g$, $\alpha_r$, $\alpha_i$ which describe the reprocessing of radiation which occurs when the stars irradiate each other. We fixed the gravity and and limb darkening coefficients using the work described in \citet{Claret2020a}, using a 4-parameter limb-darkening law \citep{Claret2000}. We estimate the Doppler beaming coefficients for the system based on \citet{Claret2020b}. 

\begin{figure*}
\includegraphics[width=1.0\textwidth]{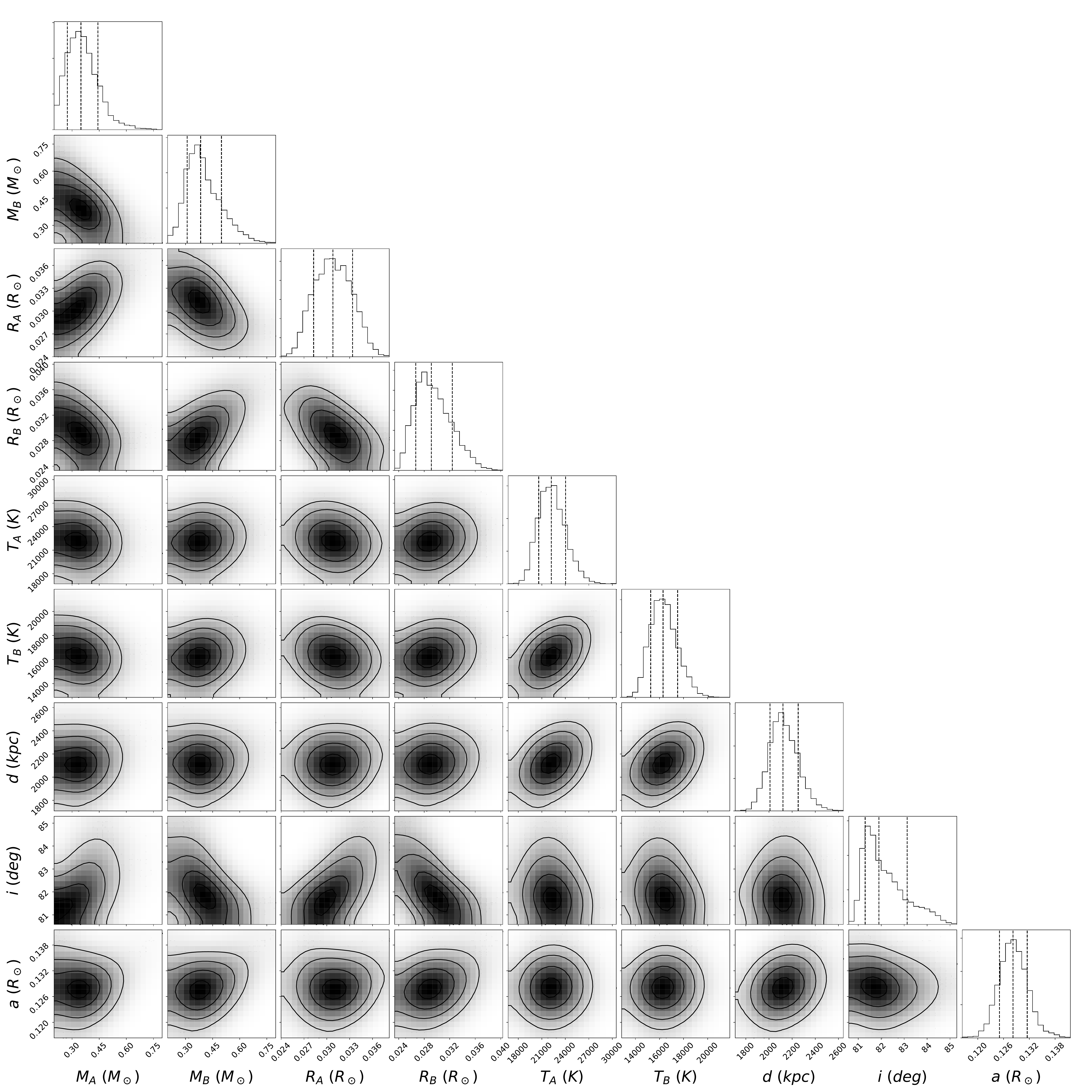}
\caption{Corner plots illustrating the covariances of quantities estimated during our combined analysis. Note that for readability, we have omitted some parameters including the time of superior conjunction, $T_0$, the orbital period, $P_b$, it's derivative, $\dot{P}_b$, and the absorption coefficients.
\label{fig:Corner}}
\end{figure*}

After constructing a likelihood function based on these free parameters, we performed our sampling using the nested sampling package, Multinest \citep{Feroz2009}. We used an evidence tolerance of $0.5$, with $1000$ live points. A final model fit to the CHIMERA $g^\prime$, $r^\prime$, and $i^\prime$ data using the parameters reported in Table \ref{tab:Parameters} is illustrated in Figure \ref{fig:CHIMERA_LCURVE}. Corner plots from this comprehensive analysis, showing the covariance between parameters, are illustrated in Figure \ref{fig:Corner}. Note that for ease of reading, we have omitted some free parameters such as $T_0$, $P_b$, $\dot{P}_b$, and the absorption coefficients. The final parameters we derived from the analysis are reported in Table \ref{tab:Parameters}. The remainder of this section discusses how we constructed our likelihood function and other details of our sampling procedure. 

\textbf{Lightcurve fit:} We simultaneously fit the CHIMERA $g^\prime$, $r^\prime$, and $i^\prime$, as well as ZTF $r$-band and $g$-band lightcurves from all nights, allowing each pass-band a free parameter representing the absorption coefficient (to model the reprocessing of radiation that arises from the stars irradiating the other, which is generally wavelength dependent). All other free parameters were the same for the lightcurve models of the three bands. Although the ZTF data has much lower SNR than the CHIMERA data, it was fit alongside the CHIMERA data because it strongly constrains the orbital period and its derivative due to its temporal baseline.

\textbf{SED Fit:} We also use the parameters we sample over to generate a synthetic SED by computing a synthetic WD model atmosphere using \citet{Tremblay2011}, with Stark broadening from \citet{Tremblay2009}, and use these synthetic spectra to compute photometry for our passbands. Because we sample over the masses and radii of the components, for each iteration, we can compute the surface gravity of both objects, and by using these in combination with the temperatures of both objects (which we also sample over), and the radii and distances to the objects, we have all the degrees of freedom needed to compute synthetic photometry for these objects. We compute the reddening for each iteration, by querying the extinction maps of \citep{Green2019}, supplying the distance of each sample to estimate the reddening for that particular iteration (which we then use to redden our synthetic photometry in order to correctly fit the SED). 

\begin{figure}
\includegraphics[width=0.5\textwidth]{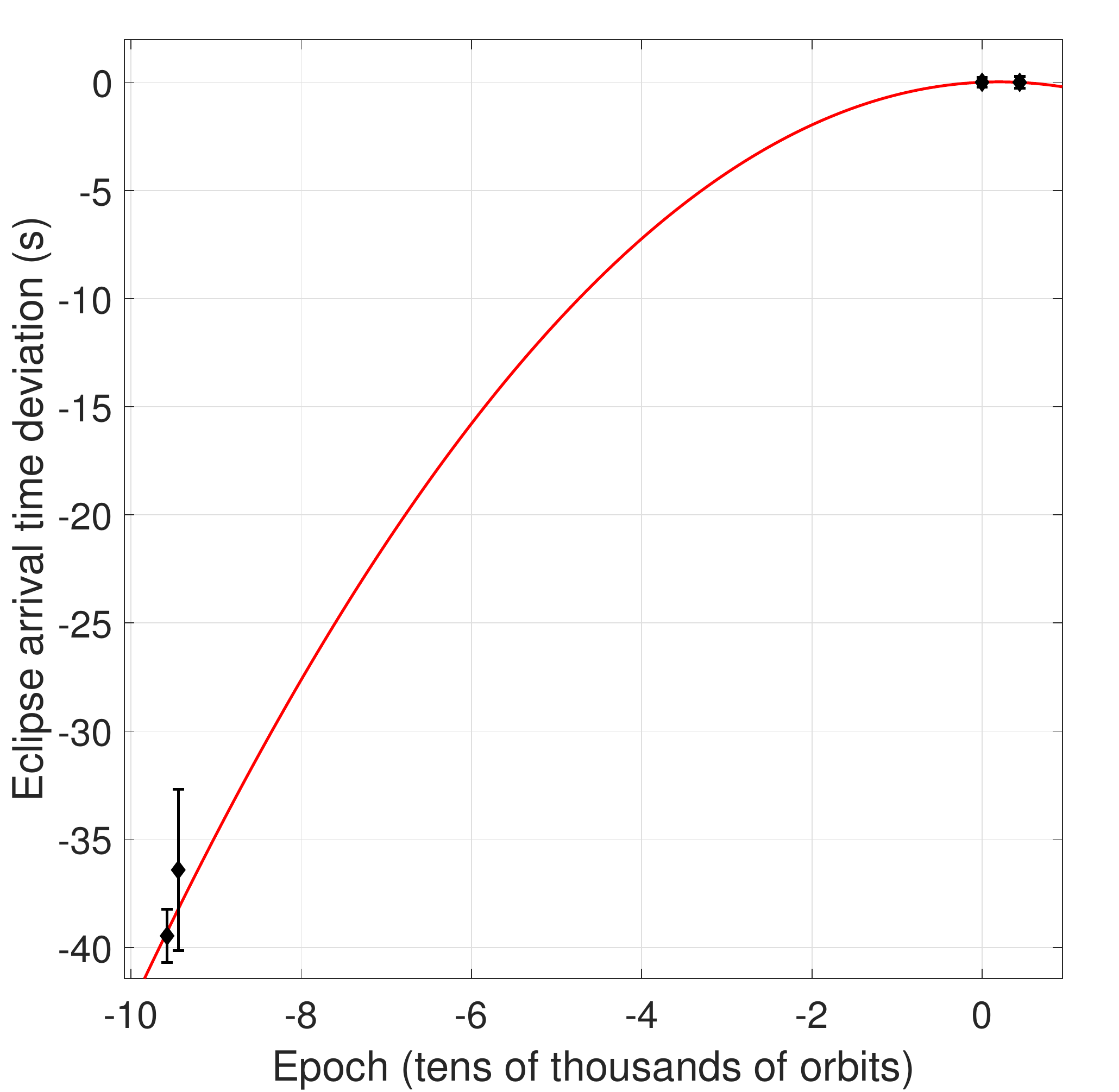}
\caption{Eclipse timing of ZTF J2243+5242, demonstrating its orbital decay. The two black diamonds on the right (with small error bars) illustrate eclipse times derived from CHIMERA data, whereas the two points at much earlier times are derived from two nights of ZTF data, each of which contains over 3 hours of continuous observations of the source. The overplotted red parabola illustrates the predicted orbital evolution based on our derived $\dot{P}_b$.
\label{fig:Decay}}
\end{figure}

\textbf{Ephemeris constraint:} We fit for the time of superior conjunction, $T_0$, which is well-constrained by the deep primary eclipse in the CHIMERA data (whose sharp ingress and egress allow for a precise measurement of the mid-eclipse time). We also fit for the orbital period $P_b$, and its derivative, $\dot{P}_b$. The latter two parameters are primarily constrained by ZTF forced photometry \citep{Yao2019}, with its two year baseline. We should be able to measure these parameters more precisely with continued monitoring of the system using high speed photometers like CHIMERA, but at present, the baseline of the CHIMERA observations is short enough that ZTF provides a far better constraint. We would like to note that unlike \citet{Burdge2019a,Burdge2019b}, we measured $\dot{P}_b$ for this system by fitting for the parameter in our lightcurve model, rather than constructing a diagram like the one shown in Figure \ref{fig:Decay} and fitting a quadratic to it. The reason for this is because there is a significant amount of ZTF data distributed throughout the last two years which contains information about the orbit between the period when the CHIMERA data was obtained and the two densely sampled ZTF nights, and thus we decided to model all the data coherently.

\textbf{Mass constraints:} In sampling over masses for the two WDs, we used a uniform prior of $0.2-0.7 \,M_\odot$ in order to speed up sampling. Our final mass estimates converged within these boundaries, indicating that we did not need to widen this prior to consider lower or higher mass solutions. The masses are primarily constrained by ellipsoidal modulation in the lightcurve, and the orbital decay of the system. 

Because we fit the ZTF lightcurves (with their long baseline) in combination with our CHIMERA data, we are able to place tight constraints on the orbital period $P_b$, and the orbital period derivative, $\dot{P}_b$. This allows us to constrain masses by assuming that the orbit is evolving according to energy loss due to gravitational-wave emission, 

\begin{equation}\label{eq:Decay}
   \dot{f}_{GW}=\frac{96}{5}\pi^{\frac{8}{3}} \left(\frac{G\mathcal{M}}{c^3}\right)^\frac{5}{3}f_{GW}^\frac{11}{3},
\end{equation} \citep{Taylor1989}, where the chirp mass is given by $\mathcal{M}=\frac{(M_AM_B)^{\frac{3}{5}}}{(M_A+M_B)^{\frac{1}{5}}}$, and the gravitational wave frequency is twice the orbital frequency, $f_{GW}=\frac{2}{P_b}$.

We use the assumption that the orbital decay is due to general relativity to place an upper bound on $\mathcal{M}$; however, it is predicted that tidal effects could significantly contribute to the evolution of a binary at these short orbital periods, and thus we estimate an additional fractional tidal contribution of approximately $7.5$ percent based on Equation 9 of \citet{Burdge2019a}, where we have taken $\kappa_A=0.12$ and $\kappa_B=0.12$, which are constants determined by the internal structure of each WD. We estimated these values based on simulations performed in \citet{Burdge2019a}, which estimated $\kappa \approx 0.11$ for the lower mass He WD in ZTF J1539+5027, and $\kappa \approx 0.14$ for the CO WD in the system (in our case, the two WDs fall between these two, and likely have a structure more similar to the He WD in ZTF J1539+5027). In any case, this approximation leads to an estimated tidal contribution of up to $7.5$ percent, and we use this constraint to place a lower bound on the chirp mass. Thus, when we sample, we sample over $\dot{P}_b$ which is fit by the lightcurves, and we also estimate a purely relativistic $\dot{P}_{b_{GW}}$ based on our masses $M_A$ and $M_B$ for that sample, and reject any solutions falling outside the range $\dot{P}_b<\dot{P}_{b_{GW}}<0.925\times \dot{P}_b$ to allow for solutions to the masses which accommodate up to a $7.5$ percent tidal contribution to the orbital evolution.

In addition to the chirp mass constraint discussed above, these masses are also constrained by the fractional amplitude of ellipsoidal variations in the lightcurve, which are given by 
\begin{equation} \label{eq:Ellipsoidal}
\frac{\Delta F_{\rm ellipsoidal}}{F}=0.15\frac{(15+u)(1+\tau)}{3-u}\left(\frac{R}{a}\right)^3q\sin^2(i),
\end{equation}
\citep{Morris1985}, where $u$ is the linear limb-darkening coefficient, and $\tau$ is the gravity darkening coefficient in the system and $q=\frac{M_B}{M_A}$ is the mass ratio of the system. In ZTF J2243+5242, the ellipsoidal variations exhibit a semi-amplitude of approximately $1.5$ percent, which is quite small compared to systems like ZTF J1539+5027. This helps constrain the masses by driving the mass ratio $q$ towards unity.

\textbf{Inclination and radius constraints:} In addition to constraining the mass ratio $q$ and the $T0$, modelling the CHIMERA lightcurve allows us to precisely estimate the inclination, $i$, and the ratio of the component radii, $R_A$ and $R_B$, with respect to the semi-major axis $a$. These constraints arise primarily from the total duration of the eclipses, and the duration of ingress/egress. Because we also sample over masses, we are able to directly constrain the semi-major axis because we know the total mass of the system, $M_A+M_B$. There is an asymmetry in the posterior distribution of the inclination, likely due to this system being on the edge of a grazing/total eclipse (it is unclear from our data whether it is flat-bottomed or not).

\textbf{Temperature constraints:} The temperatures of the two white dwarfs are constrained by an interplay of modelling the lightcurves and fitting the Pan-STARRS1 and Swift UVM2 photometry. This is because the ratio of eclipse depths in the lightcurve places stringent constraints on the surface brightness ratio and therefore the temperature ratio, whereas the SED sets the overall temperature scale. We wish to note that because of the high temperature of the objects $T>15000\, \rm K$, most of their flux is found in the ultraviolet, and thus the Swift UVM2 photometric measurement dominates this estimate, and is highly sensitive to the assumed reddening. Our solution for the temperatures is lower than that inferred from the spectroscopic modelling by about $2 \sigma$, likely due to the uncertainties in reddening. In any case, we wish to emphasize that the spectroscopic and SED temperature estimates differ by $<20$ percent, and both estimates still give a similar physical picture of the system and its evolutionary history. We wish to note that LCURVE uses a monochromatic blackbody approximation to estimate the temperature ratio of the two components based on the surface brightness ratio of the two components. By comparing with atmospheric models, we determined that the correction to the temperatures due to atmospheric effects is on the order of $\approx100\rm \, K$, which is small compared to our uncertainties.

\textbf{Distance constraints:} The distance is primarily constrained by the fit to the SED, since the overall flux contribution of each WD to the SED photometry depends only on $\frac{R}{d}$. The lightcurve fit is not directly sensitive to the distance, but it does constrain the ratio of the radii and temperature of the two WDs used in computing the synthetic photometery, as discussed above. The distance we estimate to the system, $d=2120^{+131}_{-115}$, is consistent with that of the nearby bright star to the northwest seen in Figure \ref{fig:PS}, which has a \emph{Gaia} parallax of $\bar{\omega}=0.479\pm0.024$\,mas; however, uncertainties in the astrometric solution are currently too large to establish an association (if associated, the objects would be separated by $\sim 15000$ AU).

 \subsection{Spectroscopic modelling}

The phase-resolved spectra revealed that ZTF J2243+5242 is a double-lined spectroscopic binary which consists of two hydrogen rich (DA) white dwarfs (see Figure \ref{fig:KeckSpectra}). Due to the limited signal-to-noise ratio of the phase-binned spectra we acquired, we were unable to use these spectra to fit for radial velocity semi-amplitudes in the system. The low SNRs of the spectra are a consequence of the faint nature of the object, the short exposure time needed to preserve temporal resolution, and large readout duty cycle of the observations.

However, even though we were unable to extract radial velocity semi-amplitudes from the spectra, we still were able to fit the spectra in order to estimate the effective temperatures of the WDs in the system. We use the synthetic DA NLTE white dwarf atmospheric models described in \citet{Tremblay2011}, with stark broadening from \citet{Tremblay2009}. We generate a composite WD spectrum by using the ratio of the radii of the two components inferred from lightcurve modelling to weight each component's flux appropriately, and also fix the ratio of the temperatures of the two components based on lightcurve modelling (as the relative depth of the eclipses constrains this quite well). We also fixed the surface gravities of both objects based on masses and radii inferred from lightcurve modelling. We use the masses estimated from the lightcurve analysis to appropriately Doppler shift the spectral components of each WD based on the phase of each spectrum.

By fitting the spectra with these model atmospheres, we estimate temperatures of $T_A=25700\pm600\, \rm K$ and $T_B=18500\pm400\, \rm K$. These estimates are more precise than those estimated from the SED alone (and slightly larger), and we report them here as a point of reference to compare the estimates from our combined analysis (see Table \ref{tab:Parameters}), which are based on the spectral energy distribution (SED) alone. We take both approaches to illustrate the feasibility of estimating temperature from just the SED, as this will be far more practical for the large number of faint WDs discovered by VRO and \emph{LISA} than attempting spectroscopic follow-up of these systems. An example of a fit of the spectrum of the object is illustrated in Figure \ref{fig:KeckSpectra}. Note that such an estimate is mainly feasible for eclipsing systems, in which the relative luminosity and radii of the two components can be constrained.

\begin{deluxetable}{cl}[htbp]
\tablenum{3}
\tablecaption{Physical parameters \label{tab:Parameters}}
\tablehead{\colhead{Quantity:} & \colhead{Measured value}}
\tablewidth{1000pt}

\startdata
$M_A$ & $0.349^{+0.093}_{-0.074}\,M_\odot$   \\
$M_B$ & $0.384^{+0.114}_{-0.074}\,M_\odot$     \\
$R_A$ & $0.0308^{+0.0026}_{-0.0025}\,R_\odot$  \\
$R_B$ & $0.0291^{+0.0032}_{-0.0024}\,R_\odot$  \\
$T_A$ & $22200^{+1800}_{-1600}\,\rm K$ (SED) $25700^{+600}_{-600}\,\rm K$ (Spect) \\
$T_B$ & $16200^{+1200}_{-1000}\,\rm K$ (SED) $18500^{+400}_{-400}\,\rm K$ (Spect) \\
\hline
\hline
$i $ & $81.88^{+1.31}_{-0.69}\, \rm deg$     \\
$a $ & $0.1282^{+0.0033}_{-0.0032}\,R_\odot$     \\
$T_0 $ & $59053.3448643^{+0.0000021}_{-0.0000023}\,\rm MBJD_{TDB}$     \\
$P_b $ & $527.934814^{+0.000043}_{-0.000043}\, \rm s$     \\
$\dot{P}_b$ & $1.66^{+0.17}_{-0.18}\times 10^{-11}\,\rm s\, s^{-1}$     \\
$d $ & $2120^{+131}_{-115}\,\rm pc$     \\
\enddata
\tablecomments{Measured component and orbital parameters for ZTF J2243+5242. All parameters are derived from a combined analysis of the spectral energy distribution and CHIMERA lightcurves, with the exception of $T_A$ and $T_B$, for which we also report estimates based on the optical spectrum of the system. The component parameters given here are the masses, $M_A$ and $M_B$, radii, $R_A$ and $R_B$, surface temperatures, $T_A$ and $T_B$. We also report the distance to the system $d$, and orbital parameters including the semi-major axis, $a$, inclination $i$, the time of superior conjuction $T_0$, the orbital period, $P_b$, and its derivative, $\dot{P}_b$. For the temperature estimates, $T_A$ and $T_B$, we give estimates both based on a spectroscopic fit (Spect), and based purely on the spectral energy distribution (SED).}
\end{deluxetable}

\section{Discussion}

\subsection{Evolutionary history}

Given the masses reported in Table \ref{tab:Parameters}, it is likely that the system consists of a pair of He WDs, though the uncertainties do allow for masses potentially consistent with either carbon-oxygen (CO) WDs, or hybrid WDs \citep{Perets2019}. If the system is indeed a pair of He WDs, one evolutionary channel from which ZTF J2243+5242 could have formed is via an episode of stable mass transfer, followed by a common envelope event. There is a tight relation between the radius of a star ascending the red giant branch, and its He core mass, and thus a close relationship between the mass of a He WD and the orbital period at which its progenitor star underwent a mass transfer event, stripping it of its envelope \citep{Rappaport1995}. \citet{Rappaport1995} estimated the relation as
\begin{equation} \label{eq:Mass_Period}
 P_{\rm orb}=1.3\times 10^5 M_{\rm WD}^{6.25}/(1+4M_{WD}^4)^{1.5} \, \rm days,
\end{equation} where $P_{\rm orb}$ is the orbital period at the start of the mass transfer event which forms the He WD, and $M_{\rm WD}$ is the mass of the remnant He WD in solar masses. This suggests that the progenitor of ZTF J2243+5242 underwent a common envelope event at an orbital period of $\approx 160 \rm \, days$, with an initial orbital period somewhat shorter due to the orbital widening that occurred during the preceding stable mass-transfer phase.

For a detached compact binary which has a measured $\dot{P}_b$, if one assumes that the system is undergoing orbital decay due to general relativity and can determine a cooling age, one can estimate the orbital period that the system exited common envelope by extrapolating the orbital evolution back in time. 

In order to estimate the orbital period at which the system might have exited common envelope, we modeled the evolution of the primary WD with MESA using its pre-computed WD model with $M=0.35 \, M_\odot$. Using this model, we estimate that the system is roughly 17 million years old, and exited the common envelope phase with an orbital period of 36 minutes. The actual age and initial orbital period could be slightly longer if diffusion and/or rotational mixing processes allow for more extended hydrogen burning or hydrogen shell-flashes \citep{Althaus2013}. Based on the models of \citet{Istrate2016}, these processes last less than ~10 Myr in a WD of this mass, so the system is very likely younger than ~30 Myr and was born at an orbital period less than an hour.  

A caveat to this calculation is that tidal heating may contribute significantly to the luminosity of the WDs in ZTF J2243+5242, and thus may impact our age estimates. From \cite{Burdge2019a}, the upper limit to the surface temperature produced by tidal heating is $T_{\rm tide} = (\pi \kappa M \dot{P}/ 2 \sigma_B P^3)^{1/4} \approx 30,000 \, {\rm K}$ for each of the WDs in ZTF J2243+5242. In a more realistic estimate for tidal heating, which accounts for the expected near spin-orbit synchronism, the tidal heating rate is reduced by roughly an order of magnitude, so that the tidal temperature would be closer to $T_{\rm tide} \sim 18,000 \, {\rm K}$. Hence, it is quite possible that the luminosity of the secondary is dominated by tidal heat. While tidal heating may contribute to the luminosity of the primary, its significantly higher temperature (despite a similar mass and radius) suggests its luminosity is dominated by normal white dwarf cooling, validating the young age estimate above. Hence, these rapidly merging systems may spend only a tiny fraction of their lives as DWDs.

\subsection{Future evolution}

ZTF J2243+5242 is undergoing rapid orbital decay. The system is currently clearly detached, with $\frac{R}{R_L}\approx\frac{2}{3}$ for both components; however, the two components will start interacting in approximately $320,000$ years, likely evolving into a direct impact accretor and bright source of X-rays like HM Cnc and V407 Vul. Based on the mass ratio of the system, mass transfer will likely be unstable \citep{Marsh2004}, and the system will merge in $<400,000$ years. After merger, the system is likely to form either an isolated hot subdwarf star, or an R Coronae Borealis star. In any case, the remnant of this merger will eventually cool to form a $\sim 0.5-0.7 \,M_\odot$ CO WD on the white dwarf cooling track, which may be rapidly rotating. Merging pairs of He WDs like ZTF J2243+5242 demonstrate that some ``normal-mass" CO WDs with $M\sim 0.6 \, M_\odot$ likely form from merger events.

\subsection{Implications for \emph{LISA} and the VRO}

As demonstrated in this work, using just photometric measurements, we were able to estimate component parameters for ZTF J2243+5242, including masses, temperatures, and radii, as well as orbital parameters such as inclination, period, orbital period decay rate, time of superior conjunction, and semi-major axis. This has major implications for the eras of the VRO \citep{Ivezic2019}, and \emph{LISA} \citep{Amaro2017}, which we discuss here.

\emph{LISA} and the VRO are both expected to significantly increase the number of known short period DWDs. The VRO is an upcoming optical southern sky synoptic survey using the Simonyi Survey Telescope, which has an effective aperture of 6.5-m, and the instrument has a field of view of 9.6 square degrees, about a quarter of ZTF's \citep{Ivezic2019}. The survey is expected to reach a 5-sigma depth of approximately $24.5$ in $r$ in a $30\,\rm s$ exposure, about 4 magnitudes fainter than ZTF. 


The VRO, with its smaller field of view, will acquire about a quarter the number of samples of ZTF in an equivalent survey time, and thus will not perform as well in recovering periodic objects at the same signal-to-noise ratio. By the time the survey does reach a comparable number of samples to ZTF after two and a half years, which should take the VRO about a decade or so, the frequency evolution of these objects will make it impossible to recover them without acceleration searches \citep{Katz2020}. The VRO could partially compensate for this by adopting two $15\rm \, s$ exposures rather than a single $30\rm \, s$ one, as this not only doubles the numbers epochs for such sources, but actually provides a crucial ingredient---high time resolution. Such exposures would be consecutive, effectively measuring both the flux and its derivative at a given time (which, for points in eclipse, is very valuable). Eclipsing DWDs such as ZTF J2243+5242 and ZTF J1539+5027 can significantly change their brightness in $<15\rm \, s$ during the ingress and egress of their primary eclipse, so such a measurement would be highly sensitive to this kind of rapid photometric  variability, greatly enhancing the facility's discovery capabilities in ultra-fast timescale optical variability. The other fundamental challenge is that the VRO will divide its exposures into many filters, complicating period finding (an important element in preparing for this survey will be to adapt a wide range of algorithms to cope with this technical challenge) \citep{VanderPlas2018}.

The VRO should contribute significantly to the discovery of low and moderate amplitude sources like ZTF J2243+5242 at $<23.0$ in $r$, where improved photometric precision can partially compensate for lack of temporal resolution compared to more densely sampled surveys such as ZTF. These binaries will be so faint that obtaining phase-resolved spectroscopy for more than a handful will be impossible without substantial time on an extremely large telescope (ELT). Our analysis of ZTF J2243+5242 gives hope that it will be feasible to characterize the photometrically variable systems among these without depending on spectroscopic follow-up. As discussed above, we were able to constrain many parameters in this system using just photometric measurements; this means that the large number of faint eclipsing binaries discoverable by the VRO (and eventually \emph{LISA}) could be characterized simply by obtaining a single high signal-to-noise lightcurve on a high speed photometer, and combining modelling of this lightcurve with a measurement of $\dot{P_b}$ using the longer baseline VRO data, or in some cases, such modelling could be possible using just the VRO lightcurves alone. Such analyses open up the possibility of characterizing a large population of such systems, and identifying properties such as masses/core compositions, which have implications for both the binary evolutionary processes which form these systems, and also the outcomes of the interactions/mergers.

\begin{figure}
\includegraphics[width=0.5\textwidth]{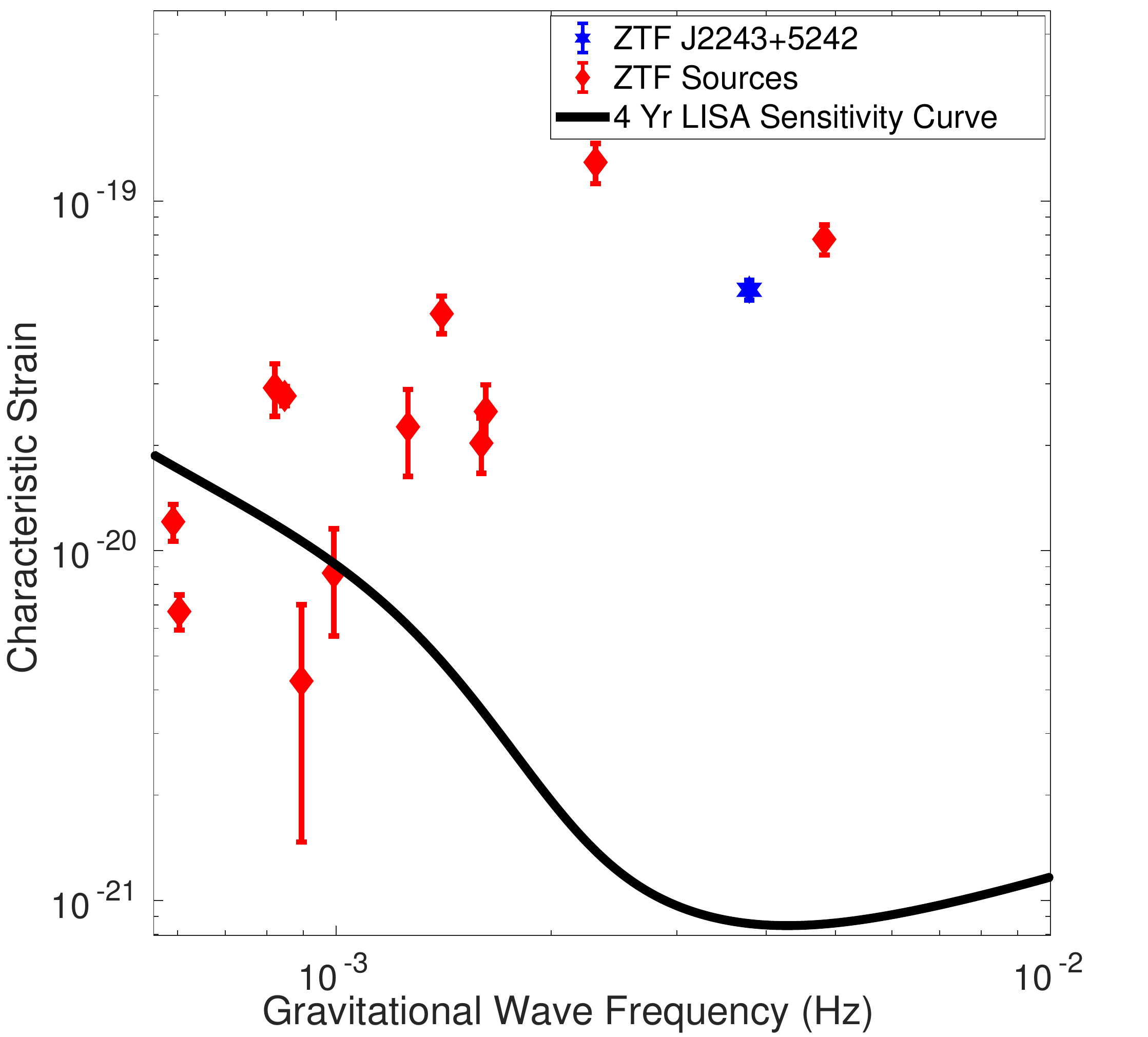}
\caption{The current ZTF sample of ultracompact binary sources reported in \citet{Burdge2020} (red diamonds). ZTF J2243+5242 is indicated as the blue six point star, the second shortest period system shown here. Because the power spectral density of \emph{LISA}'s sensitivity curve is expected to increase substantially at a frequency corresponding to around $15\,\rm min$, sources below this period, like ZTF J0538+1953, ZTF J2243+5242, and ZTF J1539+5027 should all be detected by \emph{LISA} with high signal-to-noise (SNR), enabling precise parameter estimation using GWs. It is worth noting that these high-SNR sources are all eclipsing binaries, making them particularly valuable astrophysical laboratories.
\label{fig:LISA}}
\end{figure}

In the era of \emph{LISA}, short orbital period systems like ZTF J2243+5242 and ZTF J1539+5027 will be particularly valuable astrophysical laboratories. Because these systems fall near the peak of \emph{LISA}'s sensitivity (see Figure \ref{fig:LISA}), they are detectable at large distances (ZTF J2243+5242 reaches an SNR of 7 in \emph{LISA} at $\approx 25 \, \rm kpc$, and ZTF J1539+5037 at $\approx 30 \, \rm kpc$). \emph{LISA}, which will be unhindered by Galactic extinction, should easily detect most of these kinds of objects in the Milky Way.

One way to prepare for \emph{LISA} is by developing ground and space-based instrumentation optimized to best characterize the optically detectable portion of its source population in an efficient manner. We hope that in this work and \citet{Burdge2019a,Burdge2019b,Burdge2020,Coughlin2020}, we have demonstrated that high speed photometers, which can obtain densely sampled high-signal to noise lightcurves with high temporal resolution will be one of the most powerful tools for such characterization. Such instruments on 10-m and 30-m class telescopes could be used to characterize binaries like ZTF J2243+5242 and ZTF J1539+5027 to 10-30 kpc--distances well matched to \emph{LISA}'s sensitivity threshold.

\section{Conclusion}

Here, we described the discovery and characterization of ZTF J2243+5242, the second eclipsing binary known with an orbital period under 10 minutes. The system is a DWD with an orbital period of just 8.8 minutes, and will be a strong \emph{LISA} gravitational-wave source. We performed a comprehensive analysis of the system using just photometric measurements, demonstrating the considerable value of photometry not just as a tool for discovering such extreme systems, but also one which can be used to precisely characterize these objects at great distances and faint apparent magnitudes. We were able to determine that ZTF J2243+5242 likely consists of two He WDs which will merge in approximately 400,000 years, with component masses of $M_A= 0.349^{+0.093}_{-0.074}\,M_\odot$ and $M_B=0.384^{+0.114}_{-0.074}\,M_\odot$.

ZTF has already significantly altered the landscape of extremely short orbital period binary systems known in the Galaxy. Current discoveries mark the beginning of a golden era for discovering these objects, a sample which will profoundly alter our understanding of compact binary evolution as we continue to discover more of them and understand both the processes which lead to their creation, and their eventual fates upon merger. 

\acknowledgments

K.B.B thanks the National Aeronautics and Space Administration and the Heising Simons Foundation for supporting his research.
M.W.C. acknowledges support from the National Science Foundation with grant number PHY-2010970. JF acknowledges support from an Innovator Grant from The Rose Hills Foundation, and the Sloan Foundation through grant FG-2018-10515. TRM was supported by the UK's Science and Technology Facilities Council through grant ST/T000406/1. 

Based on observations obtained with the Samuel Oschin Telescope 48-inch and the 60-inch Telescope at the Palomar Observatory as part of the Zwicky Transient Facility project. ZTF is supported by the National Science Foundation under Grant No. AST-1440341 and a collaboration including Caltech, IPAC, the Weizmann Institute for Science, the Oskar Klein Center at Stockholm University, the University of Maryland, the University of Washington, Deutsches Elektronen-Synchrotron and Humboldt University, Los Alamos National Laboratories, the TANGO Consortium of Taiwan, the University of Wisconsin at Milwaukee, and Lawrence Berkeley National Laboratories. Operations are conducted by COO, IPAC, and UW.

Some of the data presented herein were obtained at the W.M. Keck Observatory, which is operated as a scientific partnership among the California Institute of Technology, the University of California and the National Aeronautics and Space Administration. The Observatory was made possible by the generous financial support of the W.M. Keck Foundation. The authors wish to recognize and acknowledge the very significant cultural role and reverence that the summit of Mauna Kea has always had within the indigenous Hawaiian community. We are most fortunate to have the opportunity to conduct observations from this mountain.

%

\vspace{5mm}
\facilities{PO:1.2m (ZTF), Hale (CHIMERA), Keck:I (LRIS), Swift (XRT and UVOT)}


\software{Aplpy \citep{aplpy}, Astropy \citep{AstropyCollaboration2013}, CUVARBASE (\url{https://github.com/johnh2o2/cuvarbase}), LCURVE \citep{Copperwheat2010}, Lpipe \citep{Perley2019}, Multinest \citep{Feroz2009}, Numpy \citep{vanderWalt2011}          }

\vspace{5mm}
\bibliography{sample63}{}
\bibliographystyle{aasjournal}



\end{document}